\documentclass[12pt]{iopart} 

\usepackage{color,graphicx}

\begin{document}

\title{Heralded single photon absorption by a single atom}   

\author{N. Piro, F. Rohde, C. Schuck, M. Almendros, J. Huwer,
J. Ghosh, A. Haase, M. Hennrich, F. Dubin, and J. Eschner}         

\address{ICFO - Institut de Ciencies Fotoniques,\\
Mediterranean Technology Park, 08860 Castelldefels (Barcelona), Spain \\
$^*$Corresponding author: juergen.eschner@icfo.es}


\date{\today}    

\begin{abstract}
The emission and absorption of single photons by single atomic particles is a fundamental
limit of matter-light interaction, manifesting its quantum mechanical nature. At the same
time, as a controlled process it is a key enabling tool for quantum technologies, such as
quantum optical information technology \cite{Cirac1997, Monroe2002} and quantum metrology
\cite{Giovannetti2004, Leibfried2004, Schmidt2005, Roos2006}. Controlling both emission
and absorption will allow implementing quantum networking scenarios \cite{Cirac1997,
Kimble2008, Luo2009, Duan2010}, where photonic communication of quantum information is
interfaced with its local processing in atoms. In studies of single-photon emission,
recent progress includes control of the shape, bandwidth, frequency, and polarization of
single-photon sources \cite{Keller2004, Wilk2007PRLv98p63601, Legero2004PRLv93p70503,
Hijlkema2007, McKeever2004, Maunz2007, Barros2009, Almendros2009}, and the demonstration
of atom-photon entanglement \cite{Blinov2004, Volz2006, Wilk2007}. Controlled
\emph{absorption} of a single photon by a single atom is much less investigated;
proposals exist but only very preliminary steps have been taken experimentally such as
detecting the attenuation and phase shift of a weak laser beam by a single atom
\cite{Tey2008, Aljunid2009}, and designing an optical system that covers a large fraction
of the full solid angle \cite{Sondermann2007, Maiwald2009, Wrigge2008}. Here we report
the interaction of single heralded photons with a single trapped atom. We find strong
correlations of the detection of a heralding photon with a change in the quantum state of
the atom marking absorption of the quantum-correlated heralded photon. In coupling a
single absorber with a quantum light source, our experiment demonstrates previously
unexplored matter-light interaction, while opening up new avenues towards photon-atom
entanglement conversion in quantum technology.
\end{abstract}

\maketitle

Single trapped atomic ions provide optimal conditions for quantum information processing,
meeting the requirements of high-fidelity state manipulation and detection schemes, as
well as controlled interaction of the quantum bits \cite{Leibfried2003, Haeffner2008,
Blatt2008}. At the same time, single photons are ideal carriers for transmitting quantum
states and distributing entanglement over long distances \cite{Ursin2007}. Establishing
quantum correlations between single atoms and single photons allows one to perform
non-local atomic quantum gates mediated by photonic degrees of freedom, a key ingredient
in quantum networks; this was demonstrated recently in experiments which created and
employed entanglement between two remotely trapped ions \cite{Moehring2007, Maunz2009,
Olmschenk2009}. These operations are based on the underlying entanglement between a
single atom and its emitted photons. A fully bi-directional atom-photon interface implies
transfer of quantum correlations also in the absorption of a photon; then entanglement
can be distributed in a network by making two distant atoms interact with an entangled
photon pair \cite{Lloyd2001, Kraus2004}. Here, we report a step towards such entanglement
transfer by observing the interaction between a single trapped ion and resonant, heralded
single photons generated by spontaneous parametric down conversion (SPDC). We show that
the time correlation shared by SPDC photon pairs is preserved in the interaction process,
i.e. that absorption of a single photon by the ion is marked by the coincident detection
of the second photon from the pair. More generally we demonstrate that dynamics of hybrid
quantum systems involving atoms and quantum light can be controlled at the most
fundamental limit of individual quantum particles.

Figure 1 displays schematically our experimental set-up. It combines a single trapped
atomic ion with a continuous-wave source of entangled photon pairs. The $^{40}$Ca$^+$ ion
is confined and laser-cooled in a linear Paul trap which is placed between two high
numerical aperture laser objectives (HALOs) \cite{Gerber2009} collecting the ion's
laser-excited fluorescence.
The photon source emits coincident, frequency-correlated photon pairs with orthogonal
polarizations in a 200~GHz spectral band centered at the frequency of the
D$_{5/2}-$P$_{3/2}$ electronic transition of $^{40}$Ca$^+$, at 854~nm wavelength. For our
experiments, the source is designed to interact resonantly with the $^{40}$Ca$^+$ ion
\cite{Haase2009, Piro2009, Schuck2010}: the photon pairs are split, and on one of them we
impose a frequency filtering to select "trigger" photons that match in frequency and
bandwidth (25~MHz) the D$_{5/2}-$P$_{3/2}$ transition. The second, unfiltered photon is
coupled to the ion through an optical fiber and one of the HALOs. As the two coincident
photons are frequency-correlated \cite{Haase2009}, absorption events shall be accompanied
by the simultaneous detection of a trigger photon that has passed the filter. The degree
of such correlation between the detection of trigger photons and the recording of
absorption events is controlled by the detuning of the filtering cavities with respect to
the atomic resonance, and by the polarization of the photons exciting the ion.

Figure 2a presents the excitation sequence used to control the photon-ion interaction.
Each period starts by a time interval during which the motion of the ion is laser-cooled.
Excitation parameters of this part, i.e. intensities and detunings of the lasers and
duration of the pulse, are optimized experimentally to ensure that the ion is well
localized and Doppler effects are negligible. Thereafter, the internal state of the ion
is prepared using an optical pumping pulse: circularly polarized laser light at 854~nm,
propagating along the quantization axis and resonant with the D$_{5/2}-$P$_{3/2}$
electronic transition, pumps the ion into one of the two outer Zeeman sub-levels of the
D$_{5/2}$ manifold, where it remains without scattering further photons. The appropriate
helicity of the pump beam allows us to prepare an incoherent superposition of the states
either with magnetic moments $m=\left\{\frac{3}{2},\frac{5}{2}\right\}$, or with
$m=\left\{-\frac{3}{2},-\frac{5}{2}\right\}$. Finally, in the detection phase of the
sequence, the ion is exposed to the unfiltered SPDC photons, whose polarization is
controlled, while photodetectors are activated (PMT and APD in Fig.~1).

Absorption events during the detection phase are signaled by the onset of fluorescence,
following the transfer of the electronic population from the D$_{5/2}$ to the P$_{3/2}$
manifold from which spontaneous decay occurs mainly towards the S$_{1/2}$ level (with
96$\%$ probability); this then leads to steady blue fluorescence induced by the laser
excitation at 397~nm and 866~nm and recorded by the PMT photodetector. Meanwhile, the
filtered trigger photons are detected on the APD photodetector. Figure~2b presents the
signal of the two photodetectors to illustrate this process. For sufficiently high time
resolution, as shown in Fig.~2c, we note that the arrival time of the first blue photon
on the PMT coincides with the detection of a trigger photon on the APD. Hence, the time
correlation initially shared by the photon pair has been transferred to the ion-photon
system in this particular absorption event. To confirm this coincidence statistically, we
compute the second-order time correlation function, $g^{(2)}(\tau)$, between detection
events on the two detectors. Figure~2d shows that it exhibits a large peak at zero time
delay showing that absorption of a single SPDC photon by the ion is marked by the
coincident detection of a trigger photon. In the data presented in Fig.~2.d we observed a
total of 1940 absorption events, at an average rate of 1.1 s$^{-1}$. These led to 175
coincidences, 20 being accidental, such that we deduce an overall $8\%$ probability for
the transfer of the temporal quantum correlation from photon pairs to ion and photon.
While the absorption rate is determined by the brightness of the SPDC photon source
\cite{Schuck2010}, the transfer efficiency is set by the efficiency of coupling into the
single-mode fiber used to excite the single ion ($\sim$60$\%$), by the quantum efficiency
of the APD detecting the filtered photons ($\sim$35$\%$), and by the transmission through
the filtering cavities ($\sim$45$\%$).

With the ion optically pumped into the outer Zeeman sub-levels before the interaction,
the rate of coincidences between absorption events and trigger photons is controlled by
the polarization of the SPDC photons exciting the ion. We varied the polarization of the
photons from right-circular to left-circular, thereby selecting excitation of transitions
between D$_{5/2}$ and P$_{3/2}$ which involve a change of the magnetic moment by $\Delta
m=-1$ or $+1$, respectively. Transitions with $\Delta m=0$ are suppressed by exciting
along the quantization axis. Fig.~3 shows the observed dependence for preparation of the
ion in one of the two possible initial states,
$m=\left\{\frac{3}{2},\frac{5}{2}\right\}$, from which $\Delta m=1$ transitions are not
possible. The expected sinusoidal variation is found with $90\pm1\%$ visibility. Such
polarization-selective absorption of a single photon is the key to transferring the
polarization degree of freedom from the photon to the ion, and is therefore the first
step towards mapping photonic quantum information, including photonic entanglement, onto
atomic qubits.

We also studied the influence of varying the transmission frequency of the filtering
cavities. It should be noted that in this measurement the ion interacts always with the
same broadband SPDC photons. Nevertheless, the frequency correlation between the trigger
photon detected on the APD and its partner photon which may be absorbed by the ion makes
the coincidence probability of these events depend on the filter frequency setting. This
is shown in Figure~4 where we display the coincidence counts, $g^{(2)}(\tau=0)$, as a
function of the center frequency of the cavity filters. The two data sets correspond to
the two different initial states of the ion, after optical pumping with $\sigma^+$ or
$\sigma^-$ light (see Fig.~2a). Coincidences are observed within the range of frequencies
expected from the convolution of filter and transition bandwidth. The two spectra are
displaced to higher and lower energies compared to the bare resonance of the
D$_{5/2}-$P$_{3/2}$ transition. This is a consequence of the energy shift that the
magnetic sub-states of the D$_{5/2}-$P$_{3/2}$ manifold experience in the applied
magnetic field. The measured splitting of 9(2)~MHz between the centers of the two spectra
agrees, within the error, with the value of 12~MHz expected for the applied magnetic
field of 5~Gauss. Like the polarization dependence, the control of the coincidence rate
through the frequency of the trigger photons is another manifestation of the transfer of
photon properties to the atom required in bidirectional photon-atom interfaces.

\textbf{Acknowledgements.} We acknowledge support by the European Commission
(SCALA, Contract No. 015714; EMALI, MRTN-CT-2006-035369), the Spanish MICINN
(QOIT, CSD2006-00019; QLIQS, FIS2005-08257; QNLP, FIS2007-66944; CMMC,
FIS2007-29999-E), and the Generalitat de Catalunya [2005SGR00189; FI-AGAUR
(C.S.)].


\newpage


\bibliography{biblio-SPDC-ION_FR}


\newpage


\begin{figure}[hb]
\begin{center}
\includegraphics[width=.75\textwidth]{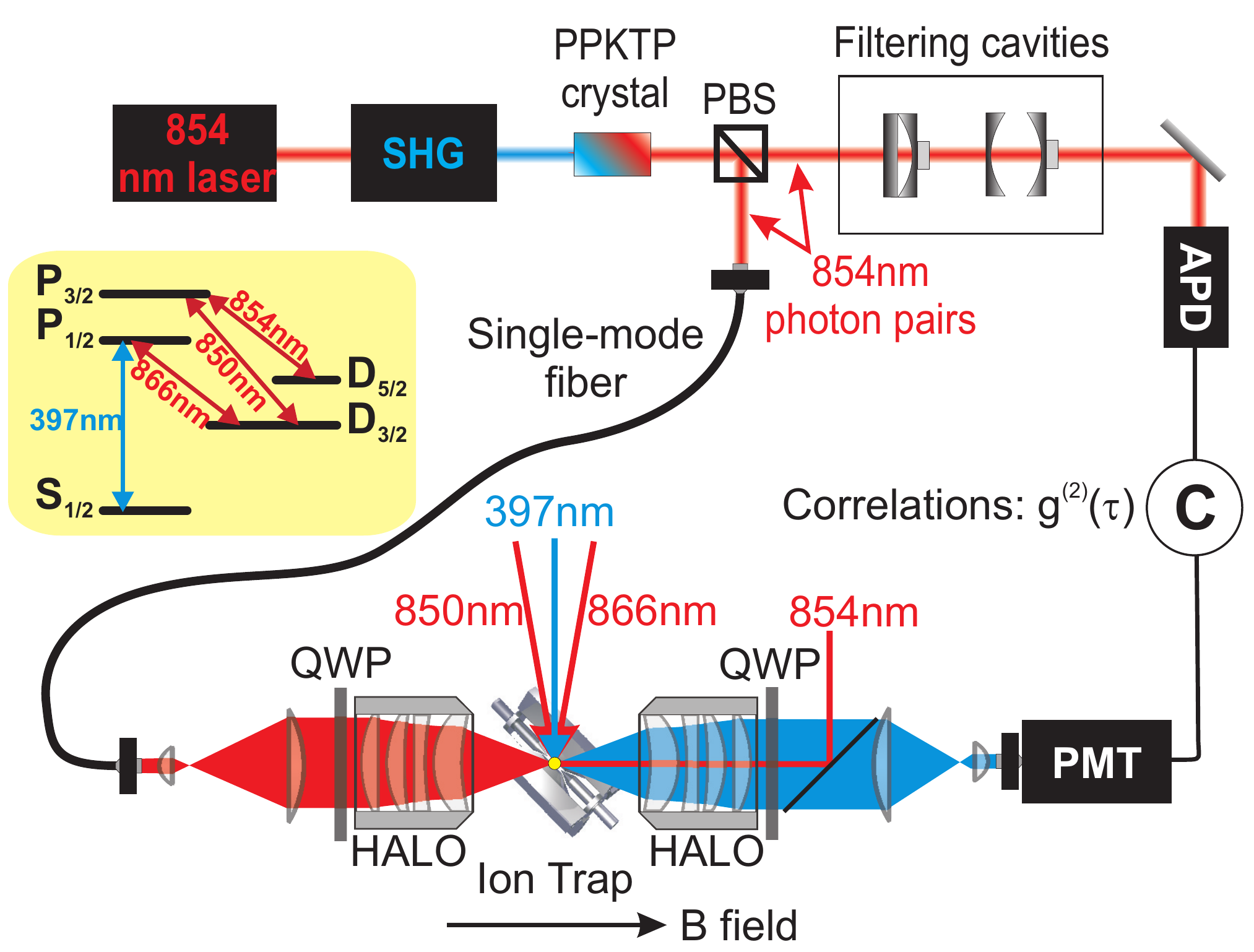}
\label{Fig:setup} \caption{\textbf{Experimental setup.}  A single $^{40}$Ca$^{+}$
ion is confined in a radio-frequency ion trap placed between two
high-numerical-aperture lenses (HALOs). The ion is laser-cooled by laser light at
397~nm and 866~nm, entering the trap from the side. Lasers at 850~nm and 854~nm
are used for state preparation; relevant atomic levels and transitions are
schematically represented in the inset. A magnetic field provides a quantization
axis along the optical axis of the HALOs. Each HALO collects about 4$\%$ of the
397~nm fluorescence photons, which are thereafter detected by a photomultiplier
tube (PMT). One HALO is also utilized to focus photons from the pair source onto
the ion.  The photon pair source is based on a narrowband, frequency-stabilized
diode laser (master laser) tuned to the D$_{5/2}-$P$_{3/2}$ transition in
$^{40}$Ca$^+$ at 854~nm. After frequency doubling (second harmonic generation,
SHG), the 427~nm light is focused into a periodically poled KTP (PPKTP) nonlinear
crystal designed to produce photon pairs around 854~nm by spontaneous parametric
down-conversion (SPDC). The crystal is operated in type-II collinear
phase-matching configuration, such that pairs of orthogonally polarized photons
are generated in a single spatial mode. A polarizing beam splitter (PBS) spatially
splits the photon pairs. One output mode is coupled through a single-mode fiber
and focussed onto the single ion through one HALO. In the second output we employ
a spectral filtering stage, consisting of two cascaded Fabry-Perot cavities, whose
transmission frequency is actively stabilized to the 854~nm master laser, while
their transmission bandwidth of 22~MHz is tailored to match the one of the 854~nm
atomic transition \cite{Haase2009, Piro2009}. Photons transmitted through the
filter are sent to an avalanche photodiode (APD), and detection events are
correlated with the arrival times of blue fluorescence photons at the PMT.
Quarterwave plates (QWP) allow controlling the polarization of the SPDC photons
and of the 854~nm laser light.}
\end{center}
\end{figure}

\newpage


\begin{figure}[tb]
\begin{center}
\includegraphics[width=.75\textwidth]{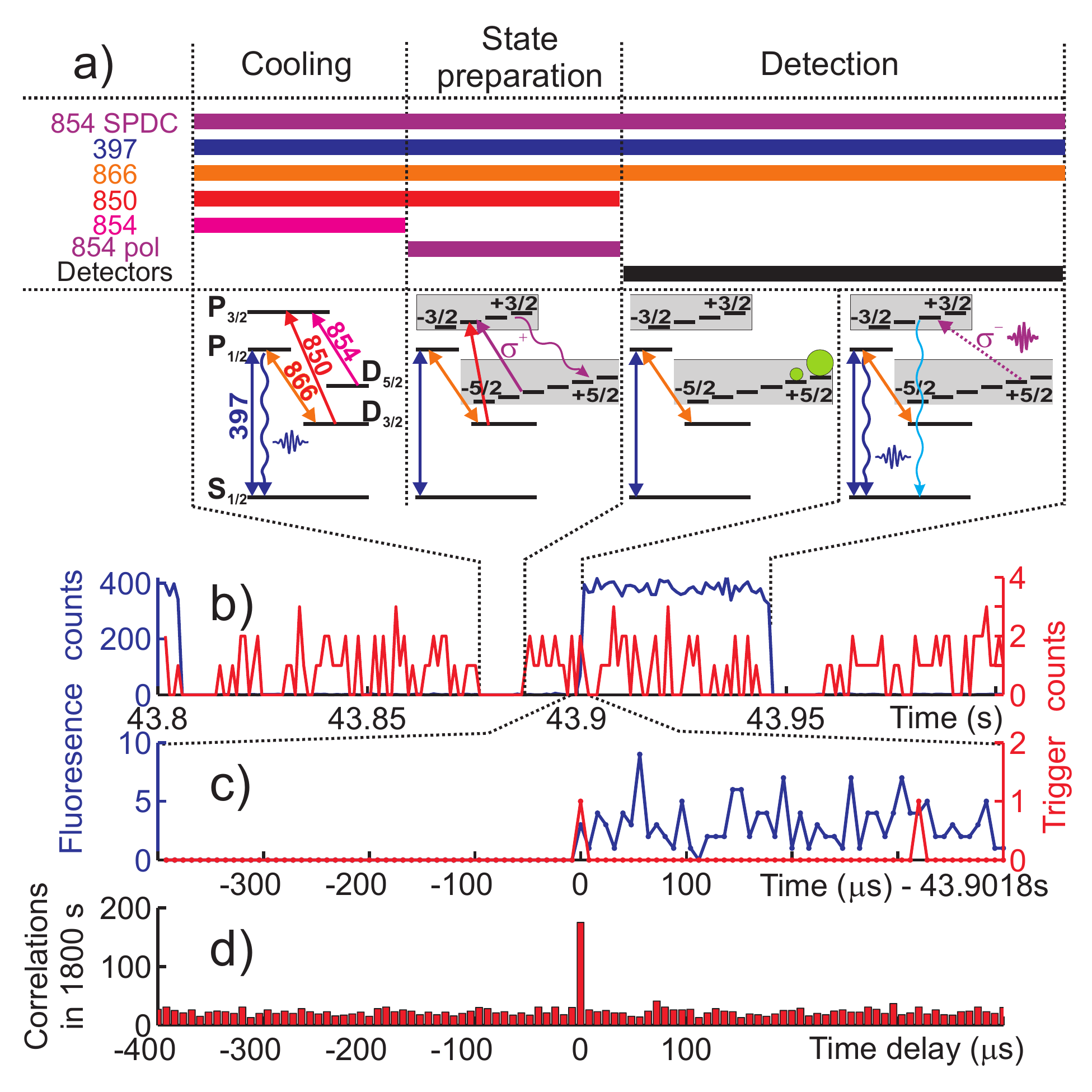}
\label{Fig:pulses} \caption{\textbf{Interaction process.} a) Periodic laser pulse
sequence used to control and observe the ion-photon interaction. It first consists
of a cooling phase lasting 5 ms followed by state preparation and detection phases
of 5 and 60 ms respectively. At the end of the preparation phase, the internal
state of the ion is deterministically initialized in sub-levels of the D$_{5/2}$
manifold with magnetic quantum numbers $m=\left\{\frac{3}{2},\frac{5}{2}\right\}$
(or $\left\{-\frac{3}{2},-\frac{5}{2}\right\}$, not shown), which couple to the
excited P$_{3/2}$ state through $\sigma^{-}$ ($\sigma^{+}$) transitions; thereby
the possible absorption of a photon in the detection phase is controlled. An
absorption event results in the onset of emission of 397~nm fluorescence by the
ion. This is illustrated in (b), where a time trace of fluorescence counts on the
PMT is displayed with 1 ms time resolution (blue line); detected trigger photons
transmitted through the filtering cavities are also shown (red). Increasing the
time resolution to 8 $\mu$s (c), we note that the photon absorption event, marked
by the first detected 397~nm photon, is coincident with the detection of its
trigger photon. The correlation between the two events is statistically verified
by a strong peak at zero time delay in the corresponding
$g^{(2)}(\tau)$-correlation function, displayed in (d) for a total recording time
of 30~min.}
\end{center}
\end{figure}

\newpage

\begin{figure}[tb]
\begin{center}
\includegraphics[width=.75\textwidth]{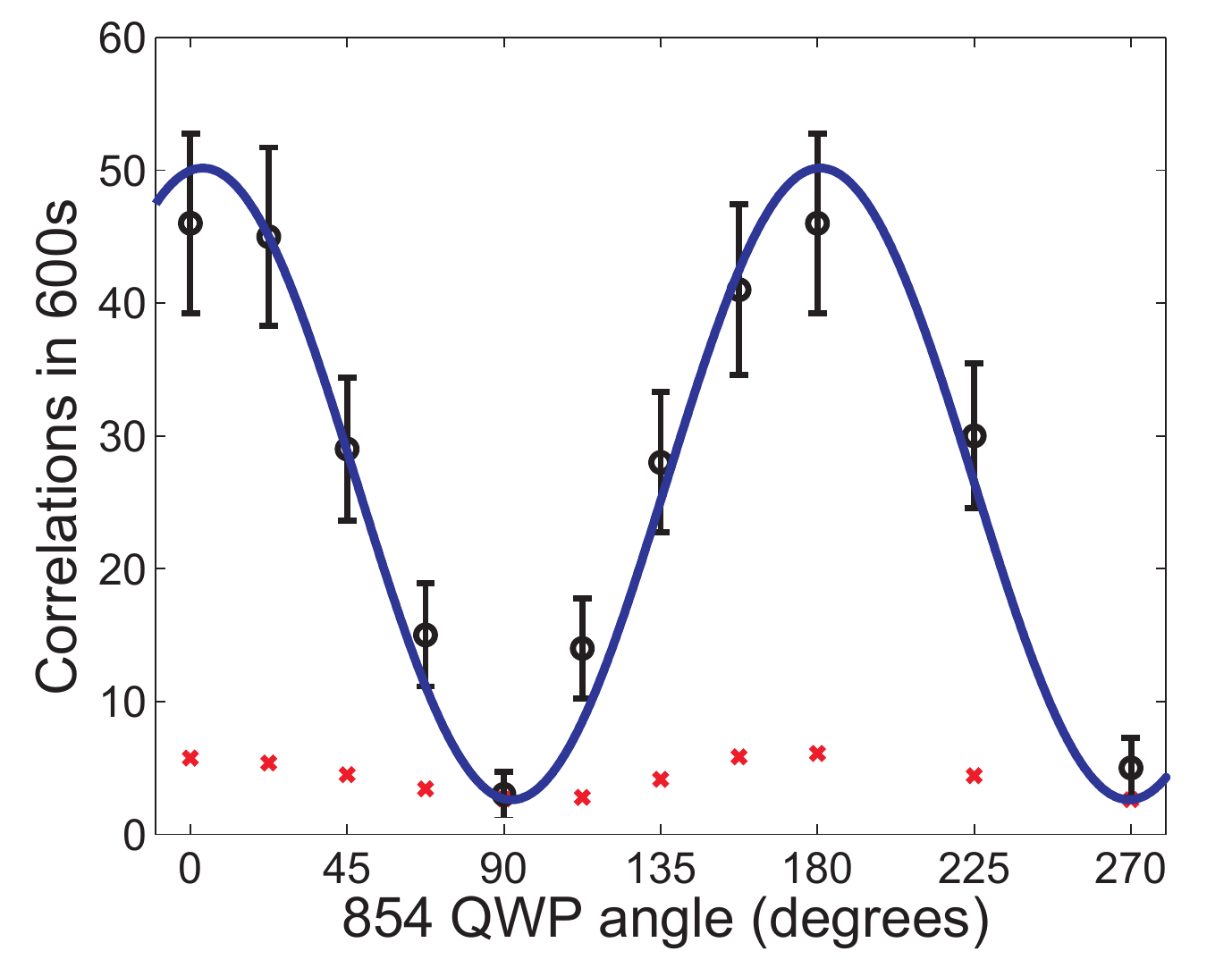}
\caption{\textbf{Dependence on photon polarization.} The ion is prepared in an incoherent
superposition of magnetic sub-levels $m=\left\{\frac{3}{2},\frac{5}{2}\right\}$ of the
D$_{5/2}$ manifold before exposure to the SPDC photons. For various settings of the
photon polarization (QWP angle), the number of coincidences between absorption events and
heralding photons is measured during 10~min, corresponding to the value of the $g^{(2)}$
correlation function at $\tau=0$ (circles). The coincidence rate reaches a maximum value
of $\approx5\textrm{min}^{-1}$ for $\sigma^{-}$ polarized photons (QWP angle of
0$^{\circ}$) and reduces to the background level (red crosses) for $\sigma^{+}$ polarized
photons (90$^{\circ}$). Displayed error bars correspond to one standard deviation
assuming Poissonian counting statistics. A visibility of $90\pm1\%$, without subtracting
accidental coincidences, is derived from a least-squares sinusoidal fit (solid curve).}
\end{center}
\end{figure}

\newpage

\begin{figure}[htb]
\begin{center}
\includegraphics[width=.75\textwidth]{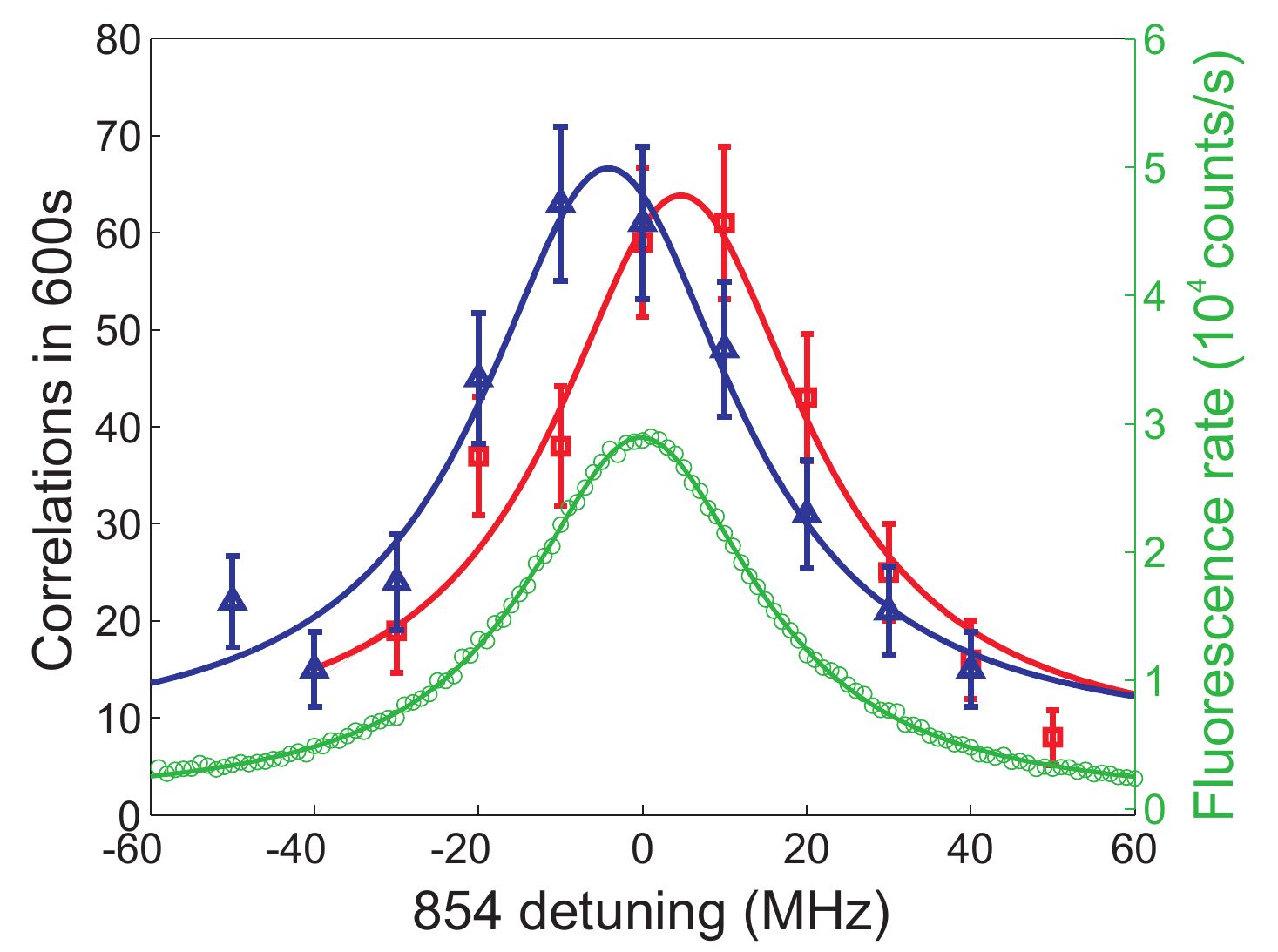}
\caption{\textbf{Correlation spectroscopy with single photons.} The rate of coincidences
between absorption events and trigger photon detection is varied by controlling the
central frequency of the filter cavities. Red squares (blue triangles) show data taken
with the 854~nm pumping laser $\sigma^{+}$ ($\sigma^{-}$) polarized and the SPDC photons
set to $\sigma^{-}$ ($\sigma^{+}$). Each data point corresponds to the value of
$g^{(2)}(\tau=0)$ obtained after 10 minutes of acquisition. Poissonian error bars are
also displayed. The center frequency of the D$_{5/2}-$P$_{3/2}$ transition is set to
0~MHz and deduced from fluorescence spectroscopy with a vertically polarized 854~nm laser
and without optical pumping (green circles). Solid lines show least-squares Lorentzian
fits to the data points.}
\end{center}
\end{figure}

\end{document}